\documentclass[twocolumn,secnumarabic,amssymb, amsmath, nobibnotes, aps, prd]{revtex4-1}

\usepackage[dvipdfmx]{graphicx}  
\usepackage[dvipdfmx]{color}
\usepackage{dcolumn}   
\usepackage{bm}        
\usepackage{amssymb}   
\usepackage{here}
\usepackage{amsmath}
\usepackage{comment}
\usepackage{color}

\usepackage{braket}
\usepackage{mathrsfs}
\usepackage{amsmath}
\usepackage{ulem}
\usepackage{fancybox}
\usepackage{xspace}
\usepackage{amsfonts}
\usepackage{multirow}
\usepackage{amsthm}
\usepackage{setspace}
\usepackage[subrefformat=parens]{subcaption}
\usepackage{ulem}
\usepackage[version=3]{mhchem}

\begin{document}

\title{Nonlinear magnon spin Nernst effect in antiferromagnets\\ and strain-tunable pure spin current}%

\author{Hiroki Kondo and Yutaka Akagi}
\affiliation{
Department of Physics, Graduate School of Science, The University of Tokyo, 7-3-1 Hongo, Tokyo 113-0033, Japan}
\email[]{kondo-hiroki290@g.ecc.u-tokyo.ac.jp}

\begin{abstract}
In this paper, we study the spin Nernst effect 
of magnons in the nonlinear response regime.
We derive the formula for the nonlinear magnon spin Nernst current by solving the Boltzmann equation and find out that it is described by an extended Berry curvature dipole of magnons.
The nonlinear magnon 
spin Nernst effect is expected to occur in various N\'eel antiferromagnetic materials even without the Dzyaloshinskii-Moriya interaction.
In particular, the nonlinear spin Nernst current in the 
honeycomb and diamond lattice antiferromagnets 
can be controlled by strain/pressure. 
\end{abstract}

\maketitle

\section{INTRODUCTION}

The Berry phase and curvature play an essential role in modern condensed matter physics; e.g., they are responsible for polarization, orbital magnetism, and various types of Hall effects~\cite{Xiao10}.
In particular, it is well-known that in the linear response regime,
the Hall effect~\cite{Klitzing80,Thouless82} and the spin Hall effect~\cite{Murakami03,Sinova15,Kane05a,Kane05b} 
are described by the integral of Berry curvature (BC).
Recently, transport phenomena have been further explored by taking into account the nonlinear response contributions. 
A remarkable study is the nonlinear Hall effect~\cite{Sodemann15}.
Even in materials with time-reversal symmetry, transverse electric current can emerge as the second-order response to an electric field~\cite{Sodemann15,Du18,Battilomo19,Son19,You18,Zhang18,Xu18,Hu20,Zhang20,Shao20,Zeng19,Yu19,Zeng20,Ma19,Kang19,Du21a,Du21b}.
It originates from a dipole moment of the BC in the crystal momentum space, 
named the Berry curvature dipole (BCD)~\cite{Sodemann15}, which appears in the systems breaking 
inversion and rotational symmetries.

Even in bosonic systems, 
it has also been clarified that the Berry phase and curvature are relevant to their physics,  
by the theoretical studies and experimental observation of the thermal Hall effect 
of magnons~\cite{Katsura10, Matsumoto11a, Fujimoto09, Shindou13a, Shindou13b, Kim16, Matsumoto14, Onose10, Ideue12, Chisnell15, Han_Lee17, Murakami_Okamoto17, Kawano19a, Li18, Hirschberger15, Lee15, Xu16, Wang20, Akagi20}, which are bosonic quasiparticles of spin waves. 
In association with the Berry curvature in magnets,
various topological magnon systems have been proposed; e.g.,
magnonic analogs of spin Hall insulators~\cite{Zyuzin16,Nakata17, Kondo19a,Mook19}, three-dimensional topological insulators~\cite{Kondo19b},  
topological crystalline insulators~\cite{Kondo21}, and 
Dirac and  Weyl semimetals~\cite{Pershoguba18, Li17, Bao18,Fransson16, Li16, Mook16, Owerre17, Su17, Liu19}, which provide a venue for the unprecedented transport phenomena in 
insulating magnetic materials.
Among them, it is worthy to note that
the magnon spin Nernst effect (SNE)~\cite{Cheng16, Zyuzin16, Nakata17, Mook19} has been 
observed in antiferromagnets (AFMs)~\cite{Shiomi17}, which makes it possible to generate the spin current 
of magnons with long coherence and promise their potential applications in spintronics~\cite{Maekawa12,Chumak15}.

Despite a number of studies on magnon systems,
representative transport phenomena, e.g.,
thermal Hall effect and SNE of magnons, have been
considered mostly in the magnets with the Dzyaloshinskii-Moriya interaction (DMI) or noncollinear spin configurations, which give rise to complex hopping matrix elements in the magnon Hamiltonian.
This is because their manifestation as a linear response requires the integration of the magnon BC over the whole Brillouin zone to be nonzero.
In order to relax the restrictions on such magnon transport phenomena,
we should explore the region beyond the linear response regime as one option.
As in the case of electronic systems, by 
investigating the magnon nonlinear response, 
nontrivial transport phenomena are expected to be discovered in magnets even without 
the DMI or noncollinear spin configuration.
So far, the nonlinear Hall effect~\cite{Mook18}, spin Seebeck effect~\cite{Takashima18}, and 
optical response~\cite{Proskurin18,Ishizuka19} of magnons have been proposed.
However, research on the nonlinear response of magnons is just beginning to emerge.

In this paper, we study the magnon second-order response to the temperature gradient by solving the Boltzmann equation and show that the nonlinear magnon spin Nernst current can be described by an extended BCD.
As shown later, the extended BCD is easily found in collinear N\'eel AFMs without the DMI, whereas the manifestation of BC, which is responsible for the linear magnon SNE and thermal Hall effect, requires 
the DMI or noncollinear spin configuration.
Therefore, it is expected that various N\'eel AFMs even without 
the DMI can exhibit the nonlinear SNE of magnons, while the magnon thermal Hall effect, as well as the linear magnon SNE, is absent.
As a demonstration, we apply the obtained formula to the strained honeycomb lattice AFM without the DMI.
A remarkable result is that the direction of the spin current can be controlled by tuning the strain.
We also investigate several other models for N\'eel AFMs
and find the presence of 
the extended BCD, which results in the nonlinear magnon SNE.

\section{EXPRESSION OF NONLINEAR SPIN NERNST CURRENT}

First, we derive the formula for the magnon spin Nernst current up to the second-order response to the temperature gradient.
We begin with the expression of the transverse magnon current in Ref.~\cite{Matsumoto11a}:
\begin{align}
J_{y}=-\frac{1}{\hbar V} \sum_{n, \bm{k}}\Omega_{n}(\bm{k})
\int_{0}^{\infty}d\epsilon \frac{\partial}{\partial x}\rho(E_{n}(\bm{k})+\epsilon,T(x)),
\label{eq:current}
\end{align}
where $E_{n}(\bm{k})$, $\Omega_{n}(\bm{k})$, $\rho(E,T(x))$ are the energy eigenvalue, BC of the $n$th band with the wave vector $\bm{k}$, and a distribution function for magnons with the energy $E$ under the temperature $T(x)$, respectively.
Here, we assume a steady state where
both ends of the system are in contact with heat baths 
at different temperatures. 
In such a case, it is known that the distribution of temperature can be written as a linear function of a position~\cite{Kreith12}.
Bearing this in mind, we henceforth assume that the temperature gradient is applied in the $x$-direction as $T(x)=T_{0}-x\nabla T$~\cite{comment_cross_terms}.
The constants $T_{0}$ and $\nabla T$ denote the average temperature and the temperature gradient, respectively.

Next, we consider the Boltzmann equation to derive the formula for the nonlinear response from Eq.~(\ref{eq:current}). 
The Boltzmann equation in the relaxation time approximation~\cite{Cercignani88,Callaway74,Mahan00} is written as follows:
\begin{widetext}
\begin{align}
&\frac{\partial}{\partial t}\rho(E_{n}(\bm{k})+\epsilon,T(x))+\dot{\bm{x}} \cdot \frac{\partial}{\partial \bm{x}} \rho(E_{n}(\bm{k})+\epsilon,T(x))
+\dot{\bm{k}} \cdot \frac{\partial}{\partial \bm{k}}
\rho(E_{n}(\bm{k})+\epsilon,T(x)) \nonumber \\
&=-\frac{\rho(E_{n}(\bm{k})+\epsilon,T(x))-\rho_{0}(E_{n}(\bm{k})+\epsilon,T(x))}{\tau},
\label{eq:Boltzmann}
\end{align}
\end{widetext}
where $\tau$ and $\rho_{0}(E,T(x))$ are the relaxation time of magnons and the equilibrium distribution function defined as $\rho_{0}(E,T(x))=[e^{E/T(x)}-1]^{-1}$, respectively.
Here, $\dot{\bm{x}}$ and $\dot{\bm{k}}$ are the time-derivatives of the position and the wave vector, respectively.
On the left-hand side of Eq.~(\ref{eq:Boltzmann}), the first and third terms drop out because we assume the steady state and the system without external field hereafter. 
Thus, Eq.~(\ref{eq:Boltzmann}) can be deformed as 
\begin{align}
&\rho(E_{n}(\bm{k})+\epsilon,T(x))\nonumber \\
&=\rho_{0}(E_{n}(\bm{k})+\epsilon,T(x))-\tau\dot{x}\frac{\partial}{\partial x}\rho_{0}(E_{n}(\bm{k})+\epsilon,T(x)).  
\label{eq:Boltzmann3}
\end{align}
Here, the velocity $\dot{x}$ is written as $(1/\hbar)\partial_{k_{x}} E_{n}(\bm{k})$.
Replacing $\partial/\partial x$ with $-\nabla T \partial/\partial T_0$ due to $T(x)=T_{0}-x\nabla T$,
we obtain the $x$-derivative of Eq.~(\ref{eq:Boltzmann3}) up to the second order in $\nabla T$ as follows:
\begin{widetext}
\begin{align}
\frac{\partial}{\partial x}\rho(E_{n}(\bm{k})+\epsilon,T(x)) 
&=-\nabla T \frac{\partial}{\partial T_{0}}\rho_{0}(E_{n}(\bm{k})+\epsilon,T_{0})+x(\nabla T)^2\frac{\partial^{2}}{\partial T_{0}^{2}}\rho_{0}(E_{n}(\bm{k})+\epsilon,T_{0})\nonumber \\
&\hspace{4mm}-\frac{\tau}{\hbar}(\nabla T)^2\frac{\partial E_{n}(\bm{k})}{\partial k_{x}}\frac{\partial^{2}}{\partial T_{0}^{2}}\rho_{0}(E_{n}(\bm{k})+\epsilon,T_{0}) +O((\nabla T)^3).
\label{eq:distr_partial}
\end{align}
\end{widetext}
The second term on the right-hand side of Eq.~(\ref{eq:distr_partial}) vanishes
in the whole space since it is an odd function of $x$.
Hereafter, we take the Boltzmann constant as $k_B=1$.
Substituting Eq.~(\ref{eq:distr_partial}) to Eq.~(\ref{eq:current}), we obtain the expression of the transverse magnon current as up to a second-order response to the temperature gradient $\nabla T$:
\begin{widetext}
\begin{align}
J_{y}
=\frac{\nabla T}{\hbar V} \sum_{n, \bm{k}}\Omega_{n}(\bm{k})
\frac{\partial}{\partial T_{0}}\int_{0}^{\infty}d\epsilon\rho_{0}(E_{n}(\bm{k})+\epsilon,T_{0}) +\frac{\tau(\nabla T)^2}{\hbar^{2} V} \sum_{n, \bm{k}}\Omega_{n}(\bm{k})\frac{\partial E_{n}(\bm{k})}{\partial k_{x}}
\frac{\partial^{2}}{\partial T_{0}^{2}}\int_{0}^{\infty}d\epsilon\rho_{0}(E_{n}(\bm{k})+\epsilon,T_{0})+O((\nabla T)^3).
\label{eq:transverse_0}
\end{align}
\end{widetext}
Here, by using the function $c_{1}(\rho_{0}):=(1+\rho_{0}){\rm ln}(1+\rho_{0})-\rho_{0}{\rm ln}\rho_{0}$, we can write down the following equation:
\begin{align}
\frac{\partial}{\partial T_{0}}\int_{0}^{\infty}d\epsilon\rho_{0}(E_{n}(\bm{k})+\epsilon,T_{0})
=c_{1}(\rho_{0}(E_{n}(\bm{k}),T_{0})).
\label{eq:c1}
\end{align}
By using Eq.~(\ref{eq:c1}) and
replacing the sum over $\bm{k}$ with the integral over the first Brillouin zone, the second term in the right-hand side of Eq.~(\ref{eq:transverse_0}) can be rewritten as follows:
\begin{align}
&\sum_{n, \bm{k}}\Omega_{n}(\bm{k})\frac{\partial E_{n}(\bm{k})}{\partial k_{x}}
\frac{\partial^{2}}{\partial T_{0}^{2}}\int_{0}^{\infty}d\epsilon\rho_{0}(E_{n}(\bm{k})+\epsilon,T_{0})\nonumber  \\
&=\sum_{n, \bm{k}}\Omega_{n}(\bm{k})\frac{\partial E_{n}(\bm{k})}{\partial k_{x}}\frac{\partial}{\partial T_{0}}c_{1}(\rho(E_{n}(\bm{k}),T_{0}))\nonumber  \\
&=-\sum_{n}\int_{\rm BZ} d^{2}k\Omega_{n}(\bm{k})\frac{\partial E_{n}(\bm{k})}{\partial k_{x}}
      \frac{E_{n}(\bm{k})}{T_{0}}
      \frac{\partial c_{1}(\rho_{0}(E_{n}(\bm{k}),T_{0}))}{\partial E_{n}(\bm{k})}    \nonumber  \\
&=-\frac{1}{T_{0}}\sum_{n}\int_{\rm BZ} d^{2}kE_{n}(\bm{k})\Omega_{n}(\bm{k})\frac{\partial }{\partial k_{x}}c_{1}(\rho_{0}(E_{n}(\bm{k}),T_{0}))\nonumber  \\
&=\frac{1}{T_{0}}\sum_{n}\int_{\rm BZ} d^{2}k
c_{1}(\rho_{0}(E_{n}(\bm{k}),T_{0}))
\frac{\partial }{\partial k_{x}}\left(E_{n}(\bm{k})\Omega_{n}(\bm{k})\right).
\label{eq:c1_2nd}
\end{align}
In the second equality of Eq.~(\ref{eq:c1_2nd}),  we replaced the $T_0$-derivative 
acting on the function $c_{1}(\rho_{0}(E_{n}(\bm{k}),T_{0}))$ with 
$(-E_{n}(\bm{k})/T_{0})\partial/\partial E_{n}(\bm{k})$
since the $T_0$-dependence is given in the form of $\rho_{0}(E_{n}(\bm{k}),T_{0})=[e^{E_{n}(\bm{k})/T_{0}}-1]^{-1}$.
By substituting Eqs.~(\ref{eq:c1}) and (\ref{eq:c1_2nd}) to Eq.~(\ref{eq:transverse_0}),
we obtain
the transverse magnon current up to the second-order of the temperature gradient $\nabla T$ 
as follows:
\begin{align}
J_{y}
&=\frac{\nabla T}{\hbar V} \sum_{n} \int_{\rm BZ} d^{2}k
c_{1}\left(\rho_{0}(E_{n}(\bm{k}),T_{0}) \right)\Omega_{n}(\bm{k})  \nonumber  \\
&+\frac{\tau(\nabla T)^2}{\hbar^{2} V T_{0}}\!\!\sum_{n}\!\!\int_{\rm BZ} \!\!\!\!d^{2}k
c_{1}(\rho_{0}(E_{n}(\bm{k}),T_{0}))
\frac{\partial }{\partial k_{x}}\!\left(E_{n}(\bm{k})\Omega_{n}(\bm{k})\right)\nonumber  \\
&+O((\nabla T)^3).
\label{eq:transversecurrent}
\end{align}
We note that the first term on the right-hand side has been obtained in Ref.~\cite{Matsumoto11a}.
The second term describes the nonlinear response in magnon systems.

In the following, we focus on the magnons in N\'eel AFMs,
assuming that the spins are aligned in the $z$-direction.
In terms of magnons, the systems have ${\mathcal{PT}}$-symmetry due to the perfect staggered magnetization; i.e.,
the magnon Hamiltonian $H(\bm{k})$ satisfies the following equation
\begin{align}
(\mathcal{PT})^{-1}\Sigma_{z}H(\bm{k}){\mathcal{PT}}=\Sigma_{z}H(\bm{k}),
\label{eq:PTsym}
\end{align}
where ${\mathcal P}$ and ${\mathcal T}$ are parity and time-reversal operators, respectively.
The matrix $\Sigma_{z}$ is defined as $\Sigma_{z}=\sigma_{z}\otimes 1_{N}$, where $\sigma_{i}$ ($i=x,$ $y,$ $z$),  $N$, and  $1_{N}$ are 
the $i$-component of the Pauli matrix,
the number of the sublattices in the unit cell, and $N$-dimensional identity matrix, respectively.
We note that in bosonic Bogoliubov-de Gennes (BdG) systems which contain pairing terms, 
we have to diagonalize not Hermitian matrix $H(\bm{k})$ but non-Hermitian matrix $\Sigma_{z}H(\bm{k})$ appearing in Eq.~(\ref{eq:PTsym})
to leave the commutation relations of boson operators unchanged~\cite{Kondo20}.
As long as the systems conserve the total spin in the $z$-direction $S^{z}$, 
we always have two degenerate magnon states related by ${\mathcal{PT}}$-operator.

We here write the eigenvectors 
and BC of magnons with the up (down) spin dipole moment as $\bm{\psi}_{n,\uparrow(\downarrow)}(\bm{k})$ 
and $\Omega_{n,\uparrow(\downarrow)}(\bm{k})$, respectively.
Since each magnon excitation carries the spin angular momentum $\pm \hbar$ in systems with conservation of the total spin along the $z$-direction,
the magnon spin Nernst current $J_{y}^{S}=\hbar (J_{y,\uparrow} - J_{y,\downarrow})$ can be written as follows: 
\begin{align}
J_{y}^{S}
&\!=\! \frac{\nabla T}{V}\! \sum_{n}\!\! \int_{\rm BZ} \!\!\!\!d^{2}k
c_{1}\!\!\left(\rho_{0}(E_{n}(\bm{k}),T_{0}) \right)\!\left(\Omega_{n,\uparrow}(\bm{k})\!-\!\Omega_{n,\downarrow}(\bm{k})\right)  \nonumber  \\
&\!+\! \frac{\tau(\nabla T)^2}{\hbar V T_{0}}\sum_{n}\int_{\rm BZ} d^{2}k
c_{1}(\rho_{0}(E_{n}(\bm{k}),T_{0})) \nonumber  \\
&\hspace{30mm}\times \frac{\partial }{\partial k_{x}}\left[E_{n}(\bm{k})\left(\Omega_{n,\uparrow}(\bm{k})\!-\!\Omega_{n,\downarrow}(\bm{k})\right)\right]\nonumber  \\
&+O((\nabla T)^3),
\label{eq:spin_current}
\end{align}
where the magnon BC is defined as $\Omega_{n,\uparrow(\downarrow)}(\bm{k})=-2 {\rm Im}[(\partial_{k_{x}}\bm{\psi}_{\uparrow(\downarrow)}(\bm{k}))^{\dagger}\Sigma_{z}(\partial_{k_{y}}\bm{\psi}_{\uparrow(\downarrow)}(\bm{k}))]$, slightly different from BC in fermionic systems due to the non-Hermiticity of the matrix $\Sigma_{z}H(\bm{k})$
in the magnon BdG systems~\cite{Kondo20}.
We remark that the spin Nernst current as a second-order response 
is described in terms of not BCD $D^{x}_{n,\uparrow(\downarrow)}(\bm{k}):=\partial_{k_{x}}\Omega_{n,\uparrow(\downarrow)}(\bm{k})$, but
the BCD-like quantity: $\bar{D}^{x}_{n,\uparrow(\downarrow)}(\bm{k}):=\partial_{k_{x}}[E_{n}(\bm{k})\Omega_{n,\uparrow (\downarrow)}(\bm{k})]$, which we call extended BCD.
One can see the structure similar to this term in the formula for the nonlinear Nernst effect of electrons~\cite{Yu19}.
We also note that the magnon thermal Hall current defined in Ref.~\cite{Matsumoto11a} is absent because
$\Omega_{n,\uparrow}(\bm{k})=-\Omega_{n,\downarrow}(\bm{k})$ holds due to ${\mathcal{PT}}$-symmetry.

\begin{figure}[H]
\centering
  \includegraphics[width=8.5cm]{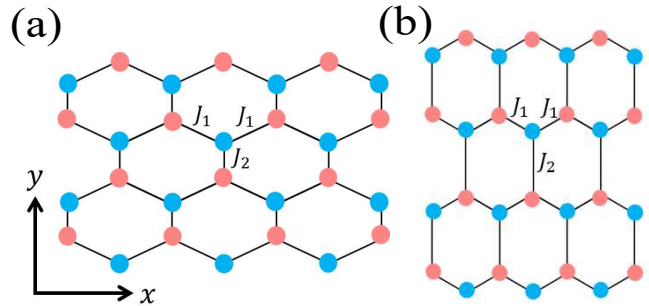}
\caption{Honeycomb lattice AFMs extended along the (a) $x$-direction and (b) $y$-direction, which correspond to (a) $J_1<J_2$ and (b) $J_1>J_2$.
Red and blue circles denote spins pointing upward and downward, respectively.
}\label{fig:honeycombstrain}
\end{figure}

\section{MODEL}

As the first model exhibiting the nonlinear magnon SNE, we consider a honeycomb lattice AFM with strain, which is illustrated in Fig.~\ref{fig:honeycombstrain}.
The Hamiltonian of the AFM is 
written as follows:
\begin{align}
\mathcal{H}=
J_{1}\sum_{\langle ij \rangle_{1}}\bm{S}_{i}\cdot\bm{S}_{j}+J_{2}\sum_{\langle ij \rangle_{2}}\bm{S}_{i}\cdot\bm{S}_{j}-\kappa\sum_{i} (S_{i}^{z})^{2},
\label{eq:Ham_AA}
\end{align}
where $\bm{S}_{i}=(S_{i}^{x},S_{i}^{y},S_{i}^{z})$ denotes the spin at site $i$.
Here, $\langle ij \rangle_{2}$ and $\langle ij \rangle_{1}$ are the nearest-neighbor vertical bonds and the other ones shown in Fig.~\ref{fig:honeycombstrain}, respectively.
The third term is an easy-axis anisotropy in the $z$-direction. By applying Holstein-Primakoff and Fourier transformations, we can obtain the magnon Hamiltonian as follows:
\begin{align}
&\mathcal{H}=\frac{1}{2}\sum_{\bm{k}}\bm{\psi}^{\dagger}(\bm{k})H(\bm{k})\bm{\psi} (\bm{k}), \nonumber \\
&H(\bm{k})=
\left(
\begin{array}{cccc}
d  &0 &0 &\gamma(\bm{k})  \\
0 &d &\gamma^{*}(\bm{k}) &0  \\
0 &\gamma(\bm{k}) &d &0  \\
\gamma^{*}(\bm{k}) &0 &0 &d  \\
\end{array} 
\right),  \nonumber \\
&\bm{\psi}^{\dagger}(\bm{k})=\left[b_{\uparrow}^{\dagger}(\bm{k}),b_{\downarrow}^{\dagger}(\bm{k}),b_{\uparrow}(-\bm{k}),b_{\downarrow}(-\bm{k})\right].\label{eq:Hk}
\end{align}
Here, $d$ and $\gamma(\bm{k}) $ are defined as $d=2J_{1}S+J_{2}S+2\kappa S$ and $\gamma(\bm{k})=2J_{1}Se^{ik_{y}/2\sqrt{3}}\cos(k_{x}/2)+J_{2}Se^{-ik_{y}/\sqrt{3}}$, respectively, where $S$ is the spin magnitude. 
The operator $b_{\uparrow(\downarrow)}(\bm{k})$ is the annihilation operator of magnons with the spin dipole moments upward (downward), i.e., magnons from spins pointing downward (upward).
The parity and time-reversal operators of this model are defined as ${\mathcal P}=1_{2}\otimes \sigma_{x}$ and ${\mathcal T}=K$, respectively.
Here, $K$ is the complex conjugation operator.
One can easily confirm that the Hamiltonian $H(\bm{k})$ satisfies the ${\mathcal{PT}}$-symmetry in Eq.~(\ref{eq:PTsym}).

Owing to the simple and typical model~(\ref{eq:Ham_AA}), we can find the candidate materials of the honeycomb AFMs.
For example, the honeycomb AFMs 
2-Cl-3,6-F$_2$-V~\cite{Okabe17} and Mn[C$_{10}$H$_6$(OH)(COO)]$_2\times$2H$_2$O~\cite{Spremo05} 
would be candidates modeled by Eq.~(\ref{eq:Ham_AA}) with $0.7<J_2/J_1<1.0$ and $J_2=2J_1$, respectively.
We note that the materials possess bond-dependences inherently.
Thus, the nonlinear magnon SNE is expected to exhibit even without the strain.

\section{NONLINEAR MAGNON SPIN NERNST EFFECT}

\begin{figure}[H]
\centering
  \includegraphics[width=8.5cm]{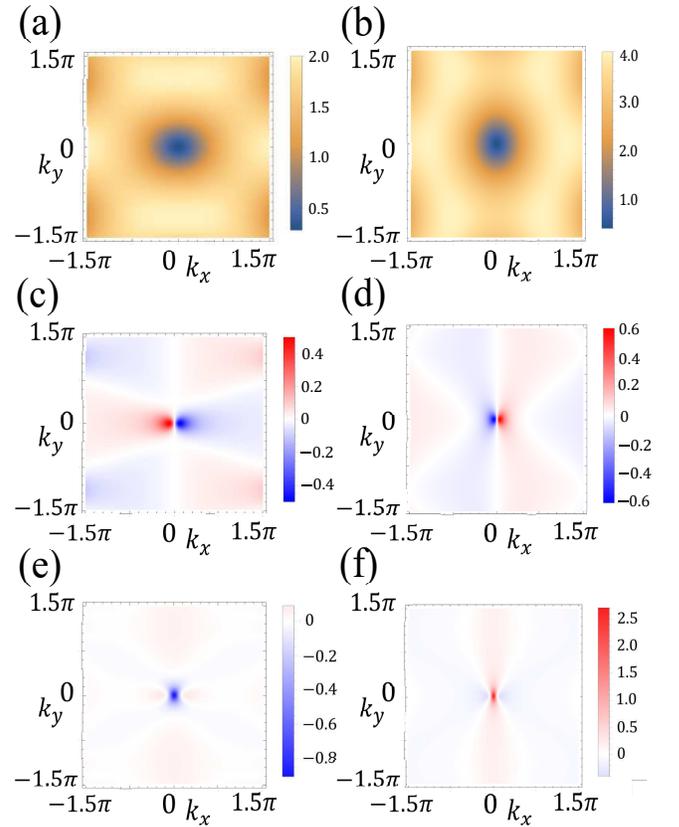}
\caption{(a), (b) Band structure $E_{\uparrow}(\bm{k})$, (c), (d) BC $\Omega_{\uparrow}(\bm{k})$, and (e), (f) extended BCD $\partial_{k_{x}}[E_{\uparrow}(\bm{k}) \Omega_{\uparrow}(\bm{k})]$ of magnons with the up spin dipole moment in the strained honeycomb lattice antiferromagnet.
The parameters in (a), (c), and (e)  are chosen to be $J_{1}S=0.5$, $J_{2}S=1.0$, and $\kappa S=0.01$, which is the case described in Fig.~\ref{fig:honeycombstrain}(a).
In (b), (d), and (f), we take the parameters to be $J_{1}S=1.5$, $J_{2}S=1.0$, and $\kappa S=0.01$, corresponding to Fig.~\ref{fig:honeycombstrain}(b).
}\label{fig:honeycomb_Berry}
\end{figure}

Figure~\ref{fig:honeycomb_Berry} shows the band structure, BC, and the extended BCD of magnons with up spin dipole moment in the strained honeycomb AFM.
Those of magnons with down spin dipole moment are determined by $E_{\downarrow}(\bm{k})=E_{\uparrow}(\bm{k})$, $\Omega_{\downarrow}(\bm{k})=-\Omega_{\uparrow}(\bm{k})$, and$\bar{D}^{x}_{\downarrow}(\bm{k})=-\bar{D}^{x}_{\uparrow}(\bm{k})$, respectively.
As shown in Figs.~\ref{fig:honeycomb_Berry}(c) and (d), BC of magnons is antisymmetric about the $\Gamma$ point.
This property corresponds to a finite extended BCD [see Figs.~\ref{fig:honeycomb_Berry}(e) and (f)]
which appears due to breaking inversion and rotational symmetries.

\begin{figure}[H]
\centering
  \includegraphics[width=8.0cm]{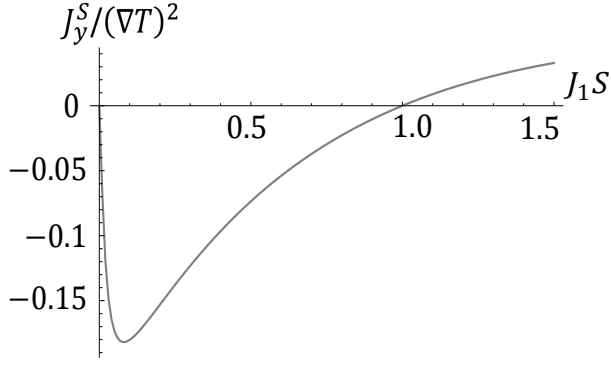}
\caption{Coefficient of the nonlinear magnon
SNE in the honeycomb AFM as a function of the coupling constant $J_{1}$.
The coupling constant $J_{2}$, the easy-axis anisotropy, and the average temperature are taken to be 
$J_{2}S=1.0$, and $\kappa S=0.01$, and $T_{0}=0.1$, respectively.
Here, we take the factor 
$\tau / (\hbar V T_{0})$ to be unity.
}\label{fig:spincurrent}
\end{figure}

Figure~\ref{fig:spincurrent} shows the coefficient of the nonlinear magnon SNE given by Eq.~(\ref{eq:spin_current}) in the honeycomb AFM as a function of the coupling constant $J_{1}$.
As expected, the coefficient becomes zero for $J_1=0$ and $J_1=J_2$ (corresponding to 
the system with the three-fold rotational symmetry restored).
It is noteworthy that spin current flows in the $+y$-direction and the $-y$-direction in the cases of $J_{1}<J_{2}$ and $J_{1}>J_{2}$, respectively.
It implies that the direction of the spin current can be controlled by tuning the strain.
Here, in which directions the transverse current is driven can be understood intuitively in terms of the balance of the coupling constants of the nearest-neighbor bonds; i.e., the transverse magnon current tends to flow in the direction of the stronger nearest-neighbor bonds
corresponding eventually to the $+y$/$-y$-direction in total.
We also note that magnon SNE in the linear response regime is absent due to the system without the DMI~\cite{LSNE_absent}.
In addition, the system does not exhibit the magnon thermal Hall effect due to $\Omega_{\downarrow}(\bm{k})=-\Omega_{\uparrow}(\bm{k})$ and $E_{\downarrow}(\bm{k})=E_{\uparrow}(\bm{k})$.

\section{ORDER ESTIMATION OF NONLINEAR SPIN NERNST CURRENT}

\begin{figure}[H]
\centering
  \includegraphics[width=5cm]{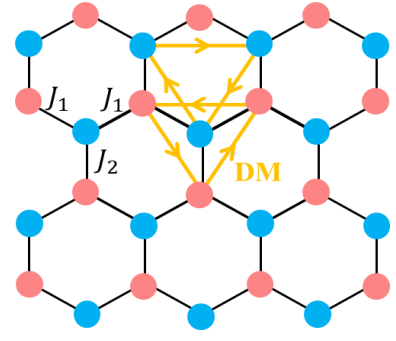}
\caption{Honeycomb lattice corresponding to the antiferromagnet MnPS$_3$ expressed by Eq.~(\ref{eq:MnPS3}).
The sign convention $\xi_{ij}=+1$ ($=-\xi_{ji}$) for $i \to j$ is 
indicated by the orange arrows.
The spin configuration is the same as in Fig.~\ref{fig:honeycombstrain}.
}\label{fig:MnPS3}
\end{figure}

Let us discuss the order of the nonlinear magnon SNE, by comparing it to the linear one observed in the honeycomb antiferromagnet MnPS$_{3}$~\cite{Shiomi17} with the DMI.
We consider the following Hamiltonian for the antiferromagnet MnPS$_3$:
\begin{align}
\mathcal{H}
&= J_{1}\sum_{\langle ij \rangle_{1}}\bm{S}_{i}\cdot\bm{S}_{j}
+J_{2}\sum_{\langle ij \rangle_{2}}\bm{S}_{i}\cdot\bm{S}_{j} \nonumber \\
&+D\sum_{\langle\langle ij \rangle\rangle}\xi_{ij}\left(\bm{S}_{i}\times\bm{S}_{j}\right)_{z}
-\kappa\sum_{i} (S_{i}^{z})^{2},
\label{eq:MnPS3}
\end{align}
where the third term is the DMI between the second nearest neighbor spins. The sign convention of 
the DMI $\xi_{ij}$ is shown in Fig.~\ref{fig:MnPS3}.
Other terms are the same as those in model (\ref{eq:Ham_AA}). 
By applying Holstein-Primakoff and Fourier transformations, we obtain the magnon Hamiltonian for model (\ref{eq:MnPS3}), which is written as in the same form as in Eq.~(\ref{eq:Hk}) with the additional term $\Delta(\bm{k})$ derived from the DMI, i.e., 
\begin{align}
&H(\bm{k})=
\left(
\begin{array}{cccc}
d - \Delta(\bm{k}) &0 &0 &\gamma(\bm{k})  \\
0 &d + \Delta(\bm{k})  &\gamma^{*}(\bm{k}) &0  \\
0 &\gamma(\bm{k}) &d + \Delta(\bm{k})  &0  \\
\gamma^{*}(\bm{k}) &0 &0 &d - \Delta(\bm{k})   \\
\end{array} 
\right),  \\
&\Delta(\bm{k}) =2DS \left[ 
-2 \sin\left(  \frac{1}{2}k_{x}  \right)     \cos\left(  \frac{\sqrt{3}}{2}k_{y}  \right)
+ \sin(k_{x})
\right].
\end{align}

\noindent

\begin{figure}[H]
\centering
\vspace{5mm}
  \includegraphics[width=8.5cm]{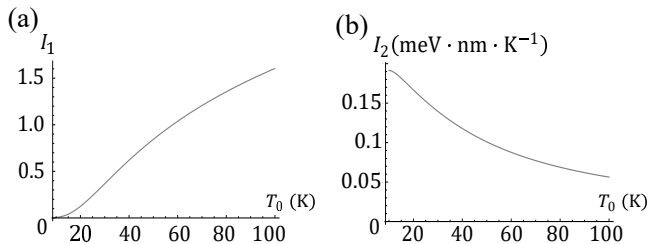}
\caption{Numerical results of (a) $I_{1}$ and (b) $I_{2}$ 
defined as Eqs.~(\ref{eq:I1})  and (\ref{eq:I2}), respectively. 
The parameters $J_{1}$ and $D$ are chosen to be (a) $J_{1}=1.54 $ ${\rm meV}$ and $D=0.36 $ ${\rm meV}$, and 
(b) $J_{1}=2.0 $ ${\rm meV}$ and $D=0$, respectively.
In both (a) and (b), the other parameters $J_{2}$, $\kappa$, and $S$ are taken as $J_{2}=1.54 $ ${\rm meV}$, $\kappa S=0.0086 $ ${\rm meV}$, and $S=5/2$, respectively.
We note that $I_{1}$ is dimensionless.
}\label{fig:compare_order}
\end{figure}

The experiment~\cite{Shiomi17} shows a good agreement with the theoretical study~\cite{Cheng16}, in which the parameters
of MnPS$_3$ are considered
as $J_{1}=J_{2}=1.54 $ ${\rm meV}$, $D=0.36$ ${\rm meV}$, $\kappa S=0.0086 $ ${\rm meV}$, and $S=5/2$~\cite{Wildes98}.
To estimate the order of linear spin Nernst current, we 
ignore the second and the third nearest-neighbor Heisenberg interactions
taken into account in Ref.~\cite{Cheng16}, whose coupling constants are less than one-fourth of the nearest-neighbor ones.
Since the antiferromagnet MnPS$_3$ 
possesses the nonnegligible DMI, the spin current observed in 
MnPS$_3$~\cite{Shiomi17} is mainly attributed to 
the linear SNE, which is described by the first term in Eq.~(\ref{eq:spin_current}). 
Here, we write the part of the integral in this term as follows:
\begin{align}
I_{1}:= \sum_{n}\int_{\rm BZ}d^{2}k
c_{1}\left(\rho_{0}(E_{n}(\bm{k}),T_{0}) \right)\left(\Omega_{n,\uparrow}(\bm{k})-\Omega_{n,\downarrow}(\bm{k})\right).
\label{eq:I1}
\end{align}
Figure~\ref{fig:compare_order}(a) shows the numerical result of 
$I_{1}$ as a function of temperature $T_0$ in model 
(\ref{eq:MnPS3}).
From the figure, the order of the spin current at $T_{0}=20 $ ${\rm K}$ 
can be written as follows:
\begin{align}
J^{S}_{\rm L}=\frac{\nabla T}{V} \times I_{1}
\sim \frac{\nabla T}{V} \times 10^{-1}. 
\label{eq:L_estimate}
\end{align}

Next, we estimate the order of nonlinear SNE in model (\ref{eq:MnPS3}) where 
the DMI is set to zero, which is equivalent to model (\ref{eq:Ham_AA}). 
We here assume that the coupling constant $J_{1}$ in Eq.~(\ref{eq:Ham_AA}) is changed as  
$J_{1}=1.54 \to 2.0$ $ {\rm meV}$. 
In such a case, i.e., for $D=0$ and $J_1 \neq J_2$, the system exhibits not the linear magnon SNE 
but the nonlinear one, which is described by the second term in Eq.~(\ref{eq:spin_current}). 
Here, we define the part of the integral of this term with the prefactor $1/T_{0}$ as $I_{2}$, i.e., 
\begin{align}
I_{2}
&:=\frac{1}{T_{0}}\sum_{n}\int_{\rm BZ} d^{2}k c_{1}(\rho_{0}(E_{n}(\bm{k}),T_{0}))
 \nonumber \\ 
&\hspace{20mm}
\times \frac{\partial }{\partial k_{x}}\left[E_{n}(\bm{k})\left(\Omega_{n,\uparrow}(\bm{k}) - \Omega_{n,\downarrow}(\bm{k})\right)\right].
\label{eq:I2}
\end{align}
The numerical result of $I_{2}$ as a function of $T_{0}$ is shown in Fig.~\ref{fig:compare_order}(b).
From this figure, we can evaluate the nonlinear spin current at $T_{0}=20 $ ${\rm K}$ as
\begin{align}
J^{S}_{\rm NL}
=\frac{\tau (\nabla T)^{2}}{\hbar V}  \times I_{2} 
\sim \frac{\tau (\nabla T)^{2}}{\hbar V}  \times \left(  10^{-1} \: {\rm meV \cdot nm \cdot K^{-1}}  \right).
\label{eq:NL_estimate}
\end{align}
Since the linear spin Nernst current 
in Eq.~(\ref{eq:L_estimate}) was observed 
with the electric voltage 
$V_{L}\sim 1$ ${\rm \mu V}$ through the inverse spin Hall effect at $T_{0}=20 $ ${\rm K}$ (see Fig.~3(c) in Ref.~\cite{Shiomi17}), we can estimate the voltage $V_{NL}$ by the 
nonlinear spin Nernst current 
in Eq.~(\ref{eq:NL_estimate}) in the following
by taking their ratio:
\begin{align}
V_{\rm NL} 
&\sim V_{\rm L}\times \frac{J^{S}_{NL}}{J^{S}_{L}} \nonumber \\
&\sim 1 \: {\rm \mu V} \times \frac{(\tau (\nabla T)^{2}/\hbar V) \times  \left(  10^{-1} \: {\rm meV \cdot nm \cdot K^{-1}}  \right) }{(\nabla T /V) \times 10^{0}}  \nonumber \\
&\sim10^{n+6}\: {\rm \mu V}.
\end{align}
Here, we assume that the lifetime of magnons and the applied temperature gradient are $\tau \sim 10^{n} $ ${\rm s}$  and $\nabla T \sim 10^{-6}$ ${\rm K\cdot nm^{-1}}$ which is estimated by the experiment in Ref.~\cite{Shiomi17}, respectively. 
In experiments, the minimum voltage detectable through the inverse Hall effect is roughly $10^{-3}$ ${\rm \mu V}$~\cite{Hirobe17}, and thus we can detect the nonlinear SNE if the magnon lifetime is $\tau \gtrsim 1$ ${\rm ns}$.
In model (\ref{eq:Ham_AA}), 
the velocity of magnons at the $\Gamma$ point is estimated as $v=(\partial/\hbar \partial k_{i}) E_{\uparrow}(\bm{k})\sim 10^{12}$ ${\rm nm \cdot s^{-1}}$.
Then, the corresponding mean free path for $V_{NL}\sim 10^{-3}$ ${\rm \mu V}$ ($\tau \sim 1$ ${\rm ns}$) is $l \sim 1$ ${\rm \mu m}$, which is achievable in magnets.

\noindent

\section{NONLINEAR MAGNON SPIN NERNST EFFECT IN VARIOUS ANTIFEROMAGNETS}

\begin{figure}[H]
\centering
  \includegraphics[width=8.5cm]{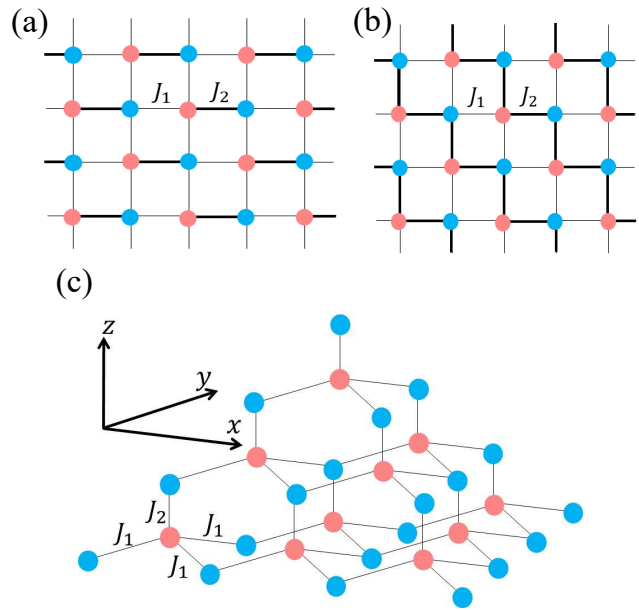}
\caption{Square lattice AFMs with (a) staggered- and (b) zigzag-bond dependences. 
The nearest-neighbor bonds $\langle ij \rangle_{1}$ and $\langle ij \rangle_{2}$ in Eq.~(\ref{eq:Ham_AA}) correspond to the ones  denoted by the thin and thick lines, respectively.
(c) Diamond lattice AFM under pressure from the $z$-direction. 
The bonds $\langle ij \rangle_{2}$ and $\langle ij \rangle_{1}$  in Eq.~(\ref{eq:Ham_AA}) correspond to the vertical and the other three ones, respectively.
Red and blue circles represent spins pointing in the $+z$ and $-z$-directions, respectively.
}\label{fig:variouslattices}
\end{figure}

In this part, we showcase the magnon extended BCD and nonlinear spin Nernst current in several AFMs; 
square lattice AFMs with  bond dependences and 
diamond lattice AFM under pressure [see Fig.~\ref{fig:variouslattices}].
The forms of the Hamiltonians in these three cases are the same as Eq.~(\ref{eq:Ham_AA}).
Figure~\ref{fig:variouslattices} shows which nearest-neighbor bonds correspond to $\langle ij \rangle_{1}$ and $\langle ij \rangle_{2}$ in the Hamiltonians.

By applying Holstein-Primakoff and Fourier transformations to Eq.~(\ref{eq:Ham_AA}), we obtain magnon Hamiltonians for these AFMs which are the same form as Eq.~(\ref{eq:Hk}) with different $d$ and $\gamma(\bm{k})$.
In the case of the square lattice AFMs with staggered-bond dependence [see Fig.~\ref{fig:variouslattices}(a)], $d$ and $\gamma(\bm{k})$ in Eq.~(\ref{eq:Hk}) is defined as  follows:
\begin{align}
&d=3J_{1}S+J_{2}S+2\kappa S, \\
&\gamma(\bm{k})=J_{1}S e^{ik_{x}/\sqrt{2}}+J_{1}S e^{ik_{y}/\sqrt{2}} \nonumber \\
&\hspace{7mm} +J_{2}S e^{-ik_{x}/\sqrt{2}}+J_{1}S e^{-ik_{y}/\sqrt{2}}.
\end{align}
Those of the square lattice AFMs with zigzag-bond dependence [see Fig.~\ref{fig:variouslattices}(b)] are given by
\begin{align}
&d=2J_{1}S+2J_{2}S+2\kappa S, \\
&\gamma(\bm{k})=J_{1}S e^{ik_{x}/\sqrt{2}}+J_{1}S e^{ik_{y}/\sqrt{2}} \nonumber \\
&\hspace{7mm} +J_{2}S e^{-ik_{x}/\sqrt{2}}+J_{2}S e^{-ik_{y}/\sqrt{2}}.
\end{align}
In the case of diamond lattice AFM [see Fig.~\ref{fig:variouslattices}(c)], $d$ and $\gamma(\bm{k})$ are written as follows:
\begin{align}
&d=3J_{1}S+J_{2}S+2\kappa S, \\
&\gamma(\bm{k})=J_{1}S e^{i\bm{k}\cdot \bm{a}_{0}}+J_{1}S e^{i\bm{k}\cdot \bm{a}_{1}}+J_{1}S e^{i\bm{k}\cdot \bm{a}_{2}}+J_{2}S e^{i\bm{k}\cdot \bm{a}_{3}},
\end{align}
where
$\bm{a}_{0}=\left(0,\frac{1}{\sqrt{3}},-\frac{1}{2\sqrt{6}}\right)$,
$\bm{a}_{1}=\left(-\frac{1}{2},-\frac{1}{2\sqrt{3}},-\frac{1}{2\sqrt{6}}\right)$,
$\bm{a}_{2}=\left(\frac{1}{2},-\frac{1}{2\sqrt{3}},-\frac{1}{2\sqrt{6}}\right)$, 
and $\bm{a}_{3}=\left(0,0,\frac{3}{4}\sqrt{\frac{2}{3}}\right)$.
One can easily make sure that these AFMs also have ${\mathcal{PT}}$-symmetry thanks to the same form of the magnon Hamiltonian as that expressed in Eq.~(\ref{eq:Hk}).

\begin{figure}[H]
\centering
  \includegraphics[width=8.5cm]{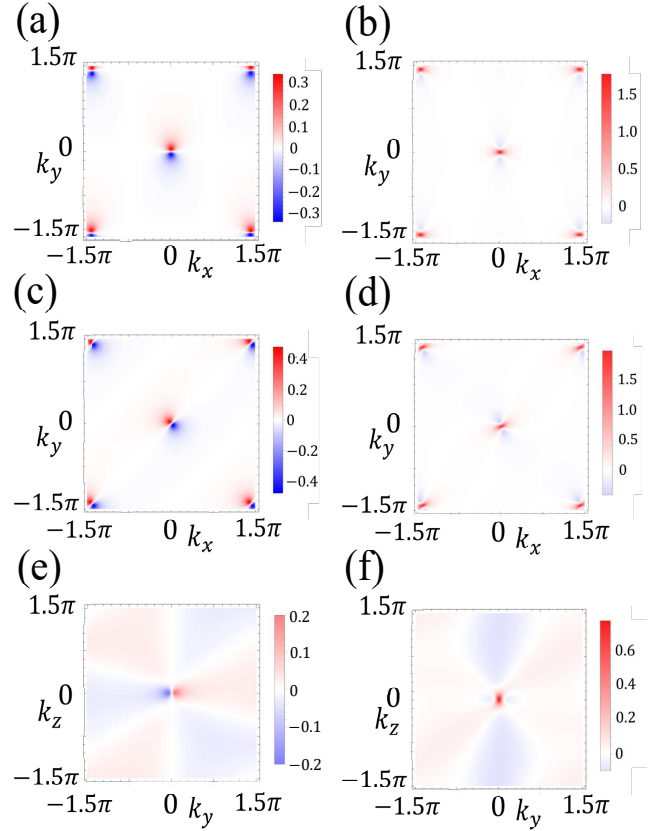}
\caption{
(a),(c) BC $\Omega_{\uparrow}(\bm{k})$ and (b),(d) extended BCD $\bar{D}_{\uparrow}^{y}(\bm{k})=\partial_{k_{y}}[E_{\uparrow}(\bm{k})\Omega_{\uparrow}(\bm{k})]$ of magnons with the up spin dipole moment in the square lattice AFM.
(a) and (b) ((c) and (d)) are results for the staggered-bond (zigzag-bond) dependence described in Fig.~\ref{fig:variouslattices}(a) (Fig.~\ref{fig:variouslattices}(b)).
Figures (e) and (f) show the $x$-component of BC $\Omega_{\uparrow}^{x}(0,k_{y},k_{z})$ and the extended BCD $\bar{D}_{\uparrow}^{xy}(0,k_{y},k_{z})$ of magnons in the diamond lattice AFM under pressure, respectively.
The parameters in these systems are chosen to be $J_{1}S=1.0$, $J_{2}S=1.2$, and $\kappa S=0.01$. 
}\label{fig:Berry_variouslattices}
\end{figure}

Figure~\ref{fig:Berry_variouslattices} shows BC and the extended BCD of magnons in these systems.
For the diamond lattice AFM, we plot the $x$-component of BC and the extended BCD defined as $\Omega_{\uparrow}^{x}(\bm{k})=-2{\rm Im}[(\partial_{k_{y}}\bm{\psi}_{\uparrow}(\bm{k}))^{\dagger}\Sigma_{z}(\partial_{k_{z}}\bm{\psi}_{\uparrow}(\bm{k}))]$ and $\bar{D}_{\uparrow}^{xy}(\bm{k})=\partial_{k_{y}}[E_{\uparrow}(\bm{k})\Omega_{\uparrow}^{x}(\bm{k})]$ in the $k_{x}=0$ plane, respectively.
They are relevant for nonlinear magnon SNE in three dimensions, as seen later [Eq.~(\ref{eq:spin_current_3D})].
As in the case of honeycomb lattice AFM, the energy eigenvalue, BC, and the extended BCD of magnons with down spin dipole moment in the square (diamond) lattice AFM are determined by $E_{\downarrow}(\bm{k})=E_{\uparrow}(\bm{k})$, $\Omega_{\downarrow}(\bm{k})=-\Omega_{\uparrow}(\bm{k})$ ($\Omega_{\downarrow}^{x}(\bm{k})=-\Omega_{\uparrow}^{x}(\bm{k})$), and $\bar{D}^{x}_{\downarrow}(\bm{k})=-\bar{D}^{x}_{\uparrow}(\bm{k})$ ($\bar{D}^{xy}_{\downarrow}(\bm{k})=-\bar{D}^{xy}_{\uparrow}(\bm{k})$), respectively.
As shown in Fig.~\ref{fig:Berry_variouslattices}, the extended BCD mainly appears around the $\Gamma$ point, which contributes to the nonlinear SNE.

\begin{figure}[H]
\centering
 \includegraphics[width=8cm]{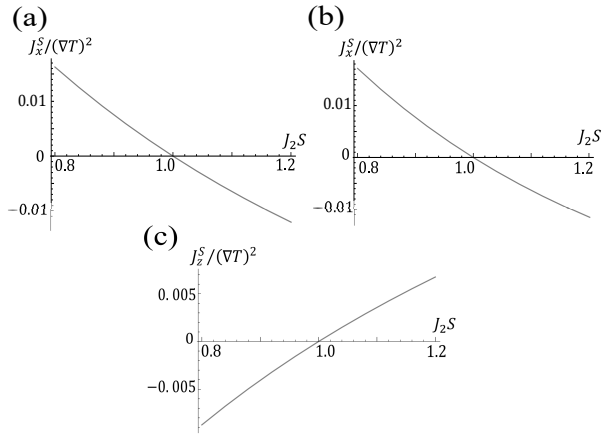}
\caption{Coefficients of nonlinear magnon SNE in the square lattice AFMs with the (a) staggered-, (b) zigzag-bond dependences, and (c) the diamond lattice AFM under pressure.
The temperature gradient is applied in the $y$-direction as $T(y)=T_{0}-y \nabla T$.
In the case of diamond lattice AFM, the spin Nernst current flows to the 
$+z$-direction ($-z$-direction) for $J_{1}<J_{2}$ ($J_{1}>J_{2}$).
We set $J_{1}S=1.0$, $\kappa S=0.01$, $T_0=0.1$ and take the factor $\tau / (\hbar V T_{0})=1$ for (a), (b), and (c).
}\label{fig:spincurrent_various}
\end{figure}

Figure~\ref{fig:spincurrent_various} shows the coefficients of nonlinear magnon SNE 
in these systems as a function of the coupling constant $J_{2}$.
For the case of the diamond lattice AFM, we generalize
the formula Eq.~(\ref{eq:spin_current}) to that in three dimensions; i.e.,
\begin{align}
J_{z}^{S}
&=\! \frac{\nabla T}{V} \!\sum_{n} \int_{\rm BZ}\!\!\! d^{3}k
c_{1}\left(\rho_{0}(E_{n}(\bm{k}),T_{0}) \right)\left(\Omega_{n,\uparrow}^{x}(\bm{k})-\Omega_{n,\downarrow}^{x}(\bm{k})\right)  \nonumber  \\
&+ \frac{\tau(\nabla T)^2}{\hbar V T_{0}}\sum_{n}\int_{\rm BZ} d^{3}k
c_{1}(\rho_{0}(E_{n}(\bm{k}),T_{0})) \nonumber  \\
&\hspace{2cm}\times \frac{\partial }{\partial k_{y}}\left(E_{n}(\bm{k})\left(\Omega_{n,\uparrow}^{x}(\bm{k})-\Omega_{n,\downarrow}^{x}(\bm{k})\right)\right)\nonumber  \\
&+O((\nabla T)^3).
\label{eq:spin_current_3D}
\end{align}
As shown in Fig.~\ref{fig:spincurrent_various},
the nonlinear SNE of magnons occurs in the square and diamond lattice AFMs while the coefficients become zero for the symmetric case, i.e., $J_1 = J_2$.
We here emphasize that Fig.~\ref{fig:spincurrent_various}(c) implies the pressure-tunable spin current can generate in the diamond lattice AFM as well as the honeycomb lattice AFM.

The quasi-two-dimensional antiferromagnet Cu(en)(H$_2$O)$_2$SO$_4$ is a candidate material of the square lattice antiferromagnet with zigzag bond dependence, in which the coupling constants are estimated as $J_{1}=10J_{2}\sim 0.30$ ${\rm meV}$~\cite{Lederova17}.
We can also find candidate materials of the diamond lattice antiferromagnets exhibiting N\'eel order, such as CoRh$_2$O$_4$~\cite{Ge18}, MnAl$_2$O$_4$~\cite{Krimmel09}, and (ET)Ag$_4$(CN)$_5$~ \cite{Shimizu19}.
In particular, (ET)Ag$_4$(CN)$_5$ is a molecular compound, and can be distorted by applying pressure without difficulty.
Thus, the pressure-tunable spin current is expected in the AFM.

\noindent

\section{SUMMARY}

In this paper, we have derived the formula for the magnon spin Nernst current as a second-order response and found that it is characterized by the extended BCD.
We then have applied the obtained formula to the strained honeycomb lattice AFMs and found out that the direction of spin current can be controlled by tuning the strain.
We have also calculated the extended BCD of magnons and confirmed that the nonlinear magnon SNE appears in the square lattice AFMs with bond dependences and the diamond lattice AFM under pressure.
Even without the DMI, the nonlinear magnon SNE is expected to be brought about in various N\'eel AFMs when the 
inversion and rotational symmetries are broken by such as strain/pressure.
Here, we note that distortion can induce the DMI, and thus the systems have the potential for exhibiting the linear SNE.
However, if materials consist of light elements, they can show mainly not the linear but nonlinear SNE because the strain-induced DMI would be negligible.

Our study reveals that the pure spin current can be generated in various N\'eel AFMs.
Owing to the simple setup, we can find a number of candidate antiferromagnetic materials exhibiting nonlinear magnon SNE; e.g., AFMs on 
a honeycomb lattice~\cite{Hiley14, Ponosov19, Okabe17, Sala21, Spremo05, Tsirlin10, Yehia10}, square lattice with zigzag bond dependence~\cite{Lederova17}, and diamond lattice~\cite{Ge18, Krimmel09, Shimizu19}.
We here emphasize that our proposal for the nonlinear magnon SNE 
provides one of 
a few possible ways to generate the spin current in materials composed of light elements such as organic materials~\cite{Naka19, Naka20} where the DMI is negligible.
Since such organic materials~\cite{Okabe17, Shimizu19} are easily deformed mechanically,
we expect to have a controllable pure spin current by strain or pressure, which expands the possibilities of applications in spintronics.

{\it Acknowledgements.}---
We acknowledge valuable discussions with Hosho Katsura.
We also thank Masahiro Sato, Hitoshi Seo, Tokuro Shimokawa for their fruitful comments. 
This work was supported by JSPS KAKENHI Grants No. JP17K14352, JP20K14411, JP20J12861,
 and JSPS Grant-in-Aid for Scientific Research on Innovative Areas
``Quantum Liquid Crystals'' (KAKENHI Grant No. JP20H05154). 
H. K. was supported by the JSPS through Program for Leading Graduate Schools (ALPS).


\begin{thebibliography}{99}
\bibitem{comment_cross_terms} 
We can also consider a more general case, i.e., $T(x,y)=T_0 -x\nabla_{x} T -y\nabla_{y} T$, where $T_0$, $\nabla_x T$, and $\nabla_y T$ are constant. 
In such a case, the component of the spin current in the direction perpendicular to $\nabla {\bar T} \equiv (\nabla_{x} T,\nabla_{y} T)$ can be formulated in the same way as Eq.~(\ref{eq:spin_current}), while  
the longitudinal component which corresponds to the nonlinear spin Seebeck effect of magnons can be expressed in the same way as in Ref.~\cite{Takashima18}. 
We note that both components of the second-order response are proportional to $|\nabla {\bar T}|^{2}=(\nabla_{x}T)^{2}+(\nabla_{y}T)^{2}$, and there is no current component proportional to $(\nabla_{x}T)(\nabla_{y}T)$.
\bibitem{LSNE_absent} It can also be understood from
BC and band structure which are antisymmetric and symmetric about the $\Gamma$ point, respectively, as shown in Figs.~\ref{fig:honeycomb_Berry}(a)-(d).
\end{thebibliography}
\end{document}